\documentclass{PoS}
\usepackage{epsfig}

\title{Bottom spectroscopy on dynamical 2+1 flavor domain wall fermion
  lattices with a relativistic heavy quark action}

\ShortTitle{Bottom spectroscopy on dynamical 2+1f DWF lattices with a RHQ action}

\author{\speaker{Min Li}\\%
        Department of Physics, Columbia University, New York, NY 10027, USA\\
        E-mail: \email{minxolee@phys.columbia.edu}}

\author{RBC and UKQCD collaborations}

\abstract{Following the successful application of the relativistic heavy
quark(RHQ) action to the charmonium and charm-strange meson spectrum,
we study here the bottom system to explore the validity of this
method in a regime with larger heavy quark momenta.  The spectrum is
calculated using the same dynamical 2+1 flavor, $24^3\times 64$ domain
wall fermion lattice configurations generated by the RBC and UKQCD
collaborations and used in the earlier charm study.  The 3 parameters
in the RHQ action are determined by matching to the experimental
bottom-strange meson masses and extrapolated to the chiral limit from
three different sea quark masses. We predict the bottomonium mass
spectrum and compare it with experiment. A theoretical estimation is
also carried out to understand the $O(a^2p^2)$ systematic errors found
in the numerical study.}

\FullConference{The XXVI International Symposium on Lattice Field Theory\\
		 July 14-19 2008\\
		 Williamsburg, Virginia, USA}

\newlength{\closercaption}
\newlength{\afterTable}
\newlength{\afterFigure}
\newlength{\closersection}
\begin{document}

\section{Introduction}
\vspace*{\closersection}

Heavy quark physics is a long standing challenge for lattice QCD due to the fact that the masses are large in current lattice units. 
Here $ma\ll 1$ is no longer true and terms containing $(ma)^n$ (with $a$ the lattice spacing) become significant. 
This problem must be addressed by using effective field theories since a direct simulation is beyond the current computation power. 
Various heavy quark effective actions were developed for different physical systems. See Refs.~\cite{Kronfeld:2003sd,Wingate:2004xa,Okamoto:2005zg,Onogi:2006km} for reviews on this topic. 

In Ref.~\cite{Li:2007en} we studied the relativistic heavy quark(RHQ) action~\cite{ElKhadra:1996mp,Aoki:2001ra,Christ:2006us,Christ:lat06} on charmonium and charm-strange systems, and found excellent accuracy for this application on charm quarks. We determined the RHQ parameters at the 
1\% level, predicted $\chi_{c0}$ and $\chi_{c1}$ masses with sub-percent accuracy, and determined the lattice scale on the $\beta=2.13$, $24^3\times 64$ 
lattice ensemble with an accuracy at least as good as that of other methods. In this proceeding we will continue to explore the three-parameter RHQ method in bottom systems, a regime with much larger heavy 
quark momenta. We start with the bottom-strange system to determine the RHQ parameters as it has smaller $a^2\vec{p}^2$ discretization errors since $p\sim\Lambda_{QCD}$, 
and then calculate the bottomonium states. To our surprise the calculations on bottomonium states have errors around 30 MeV compared to the
experiment. An attempt to theoretically estimate the $O(a^2\vec{p}^2)$ errors expected in this numerical study is also carried out at the end.

We start by briefly introducing the framework we have been using before.  The lattice form of the action, following the
formulation proposed in ~\cite{Christ:2006us,Christ:lat06}, can be written as:
\begin{equation}
  S = \sum_{n,n'} \overline{\Psi}_n\left\{ m_0 + \gamma_0D_0 -\frac{1}{2}aD^2_0 + \zeta\left[ \vec{\gamma}\cdot\vec{D}-\frac{1}{2}a(\vec{D})^2\right] 
  -a\sum_{\mu\nu} \frac{i}{4}c_P\sigma_{\mu\nu}F_{\mu\nu}\right\}_{nn'} \Psi_{n'}
\end{equation}
In heavy quark system, the temporal covariant derivative $D_0$ is on the order of $ma$ but the spatial derivatives $D_i$ are of order 
$\Lambda_{QCD}a$ or $\alpha_sma$ depending on the system under investigation. We found that only three free parameters are needed in the 
action: 
$m_0$, $c_P$ and $\zeta$. If the parameters are correctly tuned, the action will have small cutoff effects: $O(\Lambda_{QCD}a)^2$ for 
heavy-light systems and $O(\alpha_sma)^2$ for heavy quarkonium. 

The lattices used in this work are the dynamical 2+1 flavor $24^3\times 64$ DWF lattice configurations generated by the RBC-UKQCD collaborations. For better statistics, we place sources at different time slices separately for each configuration. Binning the data does not give any significant change in the size of the errors, indicating the auto-correlation of the lattice configurations is negligible. 

\begin{table}[ht]
  \centering
  \begin{tabular}{cccccc}
    \hline
    \hline
    volume & $L_s$ & ($m_{sea},m_s$) &  Traj(step)& \# of configs & sources\\
    \hline
    $24^3 \times 64$  & 16 & (0.005,0.04) & 900-6880(20) & 300    & 0,32\\
    $24^3 \times 64$  & 16 & (0.01,0.04)  & 1460-5060(40) & 91    & 0,16,32,48\\
    $24^3 \times 64$  & 16 & (0.02,0.04)  & 1885-3605(20) & 87    & 0,16,32,48\\
    \hline
  \end{tabular}
  \label{tab:lattices}
  %\caption
  %{The sources are placed at time slice 0, 16, 32 and 48 for $m_{sea}=0.01/0.02$ ensembles, and at time slices 0 and 32 for $m_{sea}=0.005$ case.}
\end{table}

\section{Determination of the RHQ action}
\vspace*{\closersection}

To determine the action in a way in which errors are controlled, the parameters are fixed by matching physical observables, 
sensitive to those parameters, to their experimental or theoretical values. 
These parameters are determined for each ensemble with three different light sea quark masses and extrapolated to the chiral limit. 
In addition to mass combinations 
of pseudo-scalar ($\eta_b$,$B_s$), vector ($\Upsilon$,$B_s^*$), scalar ($\chi_{b0}$) and axial-vector ($\chi_{b1}$) mesons in heavy-heavy (hh) and heavy-strange (hs) 
systems~\cite{Lin:2006ur}, we also calculate the tensor state ($h_b$) for heavy-heavy system. Specifically we compute: 
\begin{itemize}
\item Spin-averaged:   $m^{hh}_{sa} = \frac{1}{4}(m_{\eta_b} + 3m_{\Upsilon})$,  
  $m^{hl}_{sa} = \frac{1}{4}(m_{B_s} + 3m_{B_s^*})$  
\item Hyperfine splitting:  $m^{hh}_{hs} = m_{\Upsilon} - m_{\eta_b}$, 
  $m^{hl}_{hs} = m_{B_s^*} - m_{B_s}$
\item Mass ratio:  $\frac{m_1}{m_2}$, where $E^2 = m_1^2 + \frac{m_1}{m_2}p^2$, $m_1$: rest mass, $m_2$: kinetic mass. 
\item Spin-orbit averaged and splitting (hh): $m^{hh}_{sos} = m_{\chi_{b1}} - m_{\chi_{b0}}$,
  $m^{hh}_{soa} = \frac{1}{4}(m_{\chi_{b0}} + 3m_{\chi_{b1}})$
\item Heavy-heavy $^1P_1$ state $h_b$
\end{itemize}
Here are some sample plots of the bottomonium($\eta_b$) and bottom-strange($B_s$) pseudo-scalar correlators. One should notice the
correlators are falling about 70 orders of magnitude for bottomonium and about 40 for bottom strange. We must be careful that
the heavy propagators are sufficiently accurate when the time is large.
\begin{figure}[ht]
  \hfill
  \begin{minipage}{0.45\textwidth}
    \hspace{-1.0cm}
    \epsfig{file=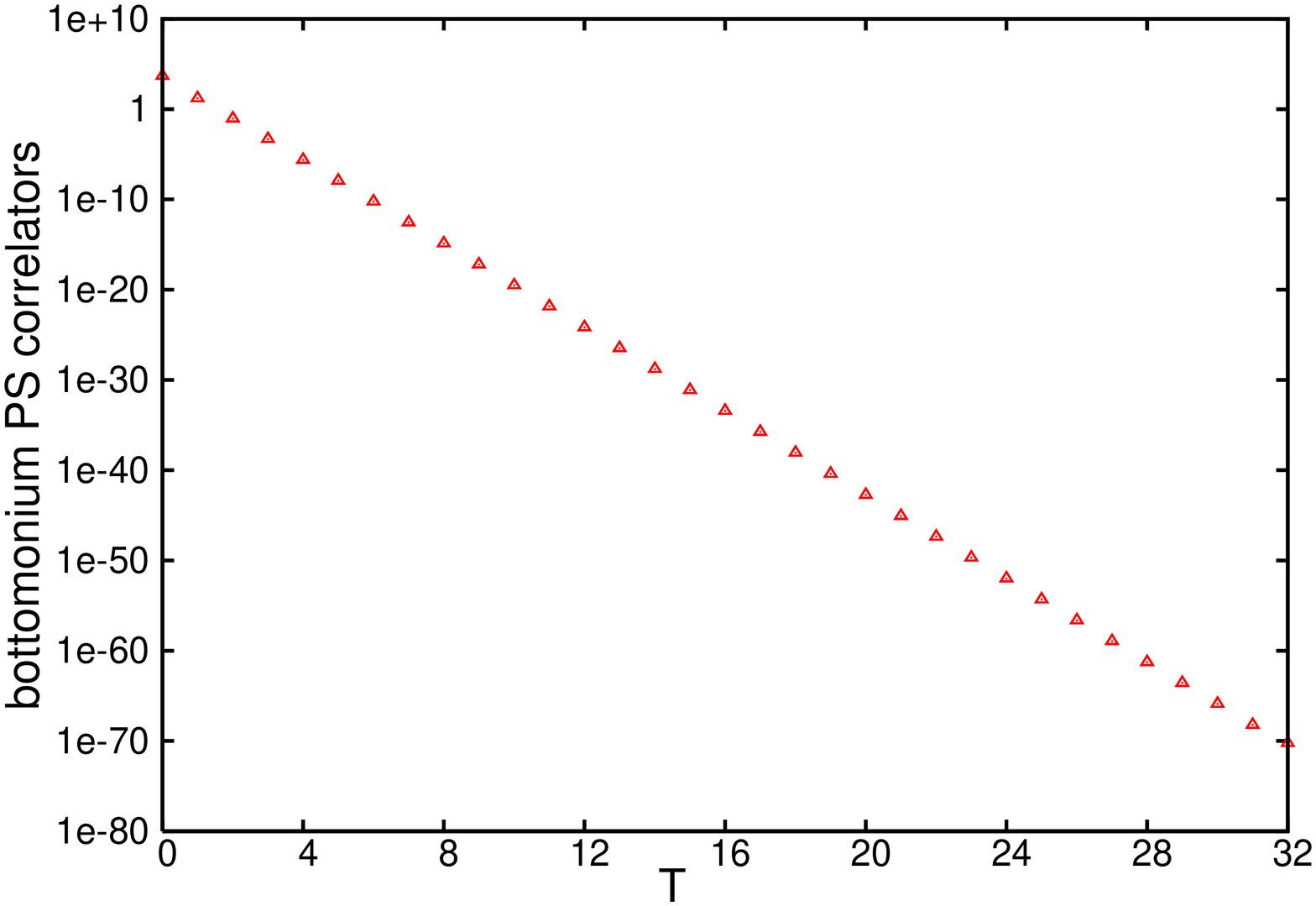,width=0.75\linewidth,angle=270}    
  \end{minipage}
  \hfill
  \begin{minipage}{0.45\textwidth}
    \hspace{-0.8cm}
    \epsfig{file=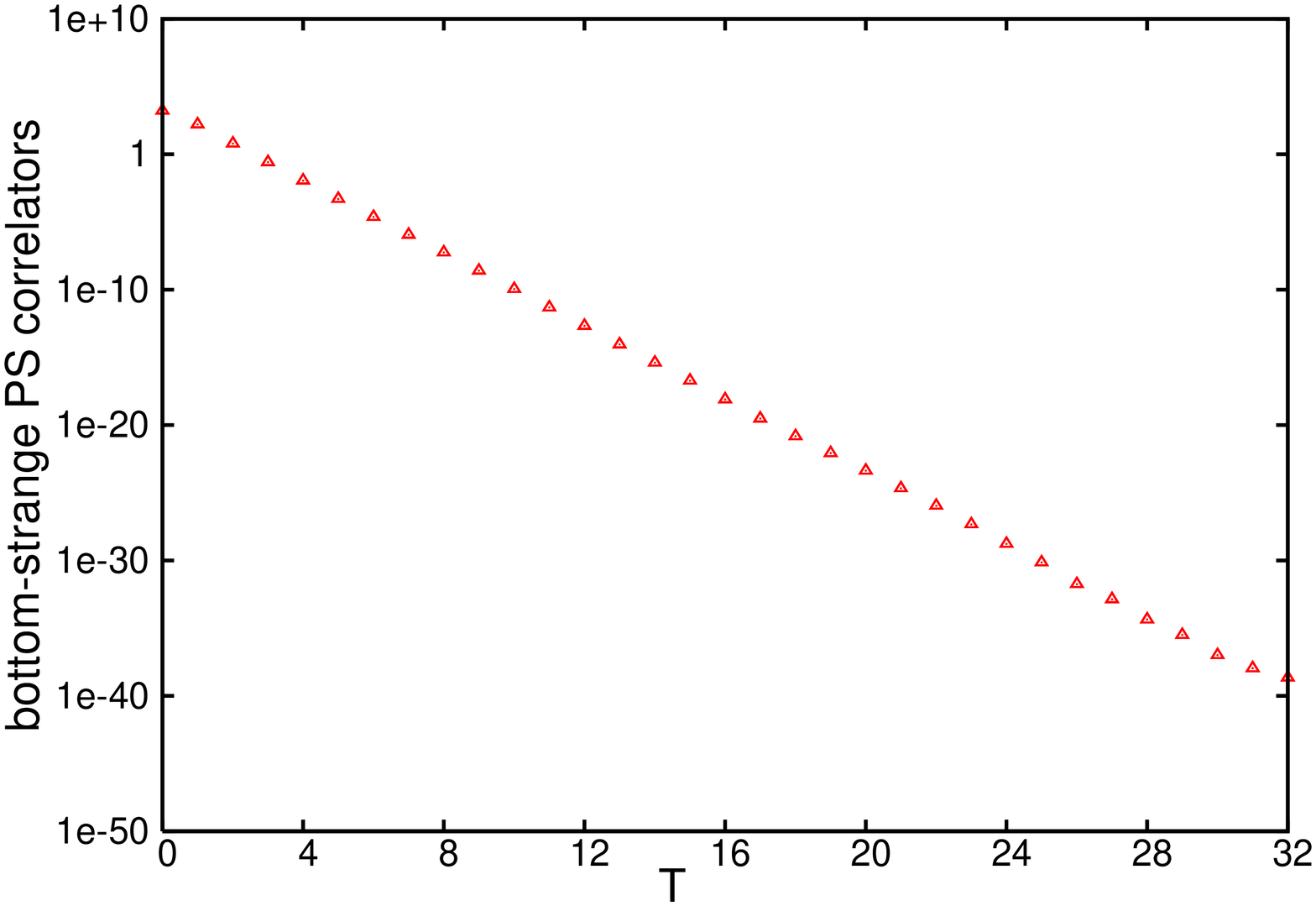,width=0.75\linewidth,angle=270}  
  \end{minipage}
  \caption{sample correlators of bottomonium and bottom-strange pseudo-scalar.}
  \label{fig:sample-crltr}
  \hfill
  \vspace*{\afterFigure}
\end{figure}

The matching method is the same as used in Ref.~\cite{Li:2007en}. We again use the linear ansatz relating the three parameters 
($X_{RHQ}$) and the corresponding measured quantities ($Y(a)$) and determine the parameters by minimizing the $\chi^2$ defined in that
 reference. The linear approximation only holds in a certain region of the parameter space, which we estimate from our earlier work using
a $16^3\times 32$ lattice. 
The linear coefficients are calculated directly using finite differences from seven point Cartesian parameter sets, which are centered at 
\{7.3,4.0,4.3\} and with extent \{0.5,1.0,0.3\}. We examine the linearity of the parameter dependence in this region and required that the
 finally matched RHQ parameters actually lie within the region examined.

%\begin{equation}
%  Y(a) = \left(
%  \begin{array}{c}
%    m_{\eta_b} a \\
%    m_{\Upsilon} a \\
%    ...\\
%    ... \\
%    m_1/m_2 
%    \end{array} 
%  \right) = J\cdot X_{RHQ}=J\cdot \left( 
%  \begin{array}{c}
%    m_0a \\
%    c_P \\
%    \zeta
%  \end{array} 
%  \right)
%  + A \nonumber
%\end{equation}
%where the quantities $Y(a)$ is known if we assume lattice spacing $a$ is known from other method
%or $a$-dependent if we treat $a$ as a free parameter to be determined. Provided we know how to get
%J matrix and A vector, we can obtain the parameters by minimize the $\chi^2$ defined as:
%\begin{equation}
%  \chi^2 = (J\cdot X_{RHQ}+A-Y(a))^TW^{-1}(J\cdot X_{RHQ}+A-Y(a)) 
%  \label{eqn:chi2}
%\end{equation}
%$W$ is the correlation matrix of the quantities estimated from the measured data. We choose to use only the diagonal part sometimes because 
%the data might be too noisy to give a not-well-behaved $W$. The $\chi^2$ is a quadratic function of vector $X_{RHQ}$ if lattice spacing $a$ 
%is known and of vector $(m_0a,c_P, \zeta, a)^T$ if $a$ is unknown, and so it is easy to minimize analytically. J and A can be calculated using 
%finite differences directly from a cartesian set, and for saving time we only get the data for the minimum (seven) number of parameter sets: 
%centered at \{7.3,4.0,4.3\} and with extend \{0.5,1.0,0.3\}. We examine the linearity of the dependence in this region and make sure the finally 
%matched RHQ parameters are actually in the region we are working.

\subsection{Numerical details}
\begin{figure}[ht]
  \hfill
  \begin{minipage}{0.45\textwidth}
    \hspace{-1.0cm}
    \epsfig{file=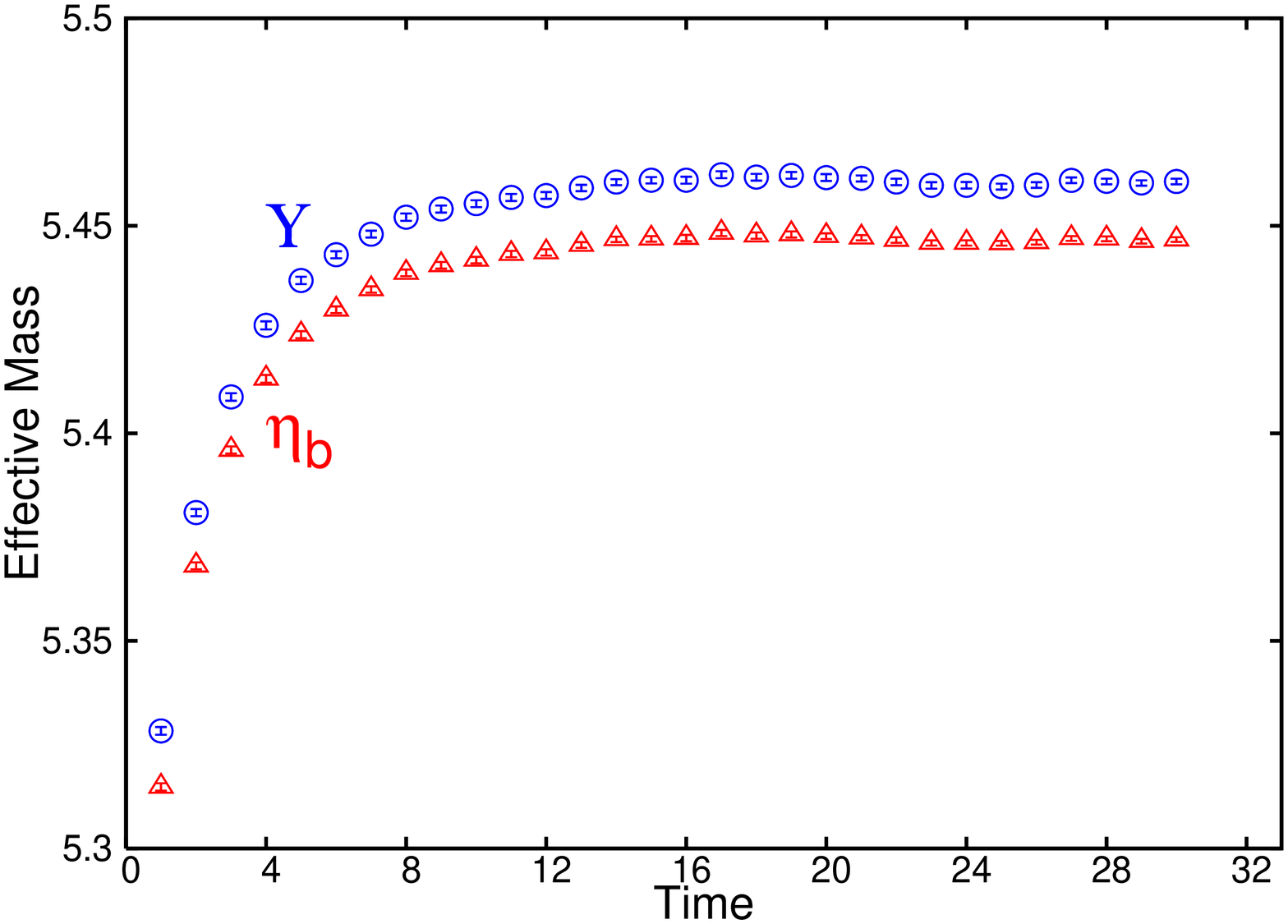,width=0.75\linewidth,angle=270}    
  \end{minipage}
  \hfill
  \begin{minipage}{0.45\textwidth}
    \hspace{-0.8cm}
    \epsfig{file=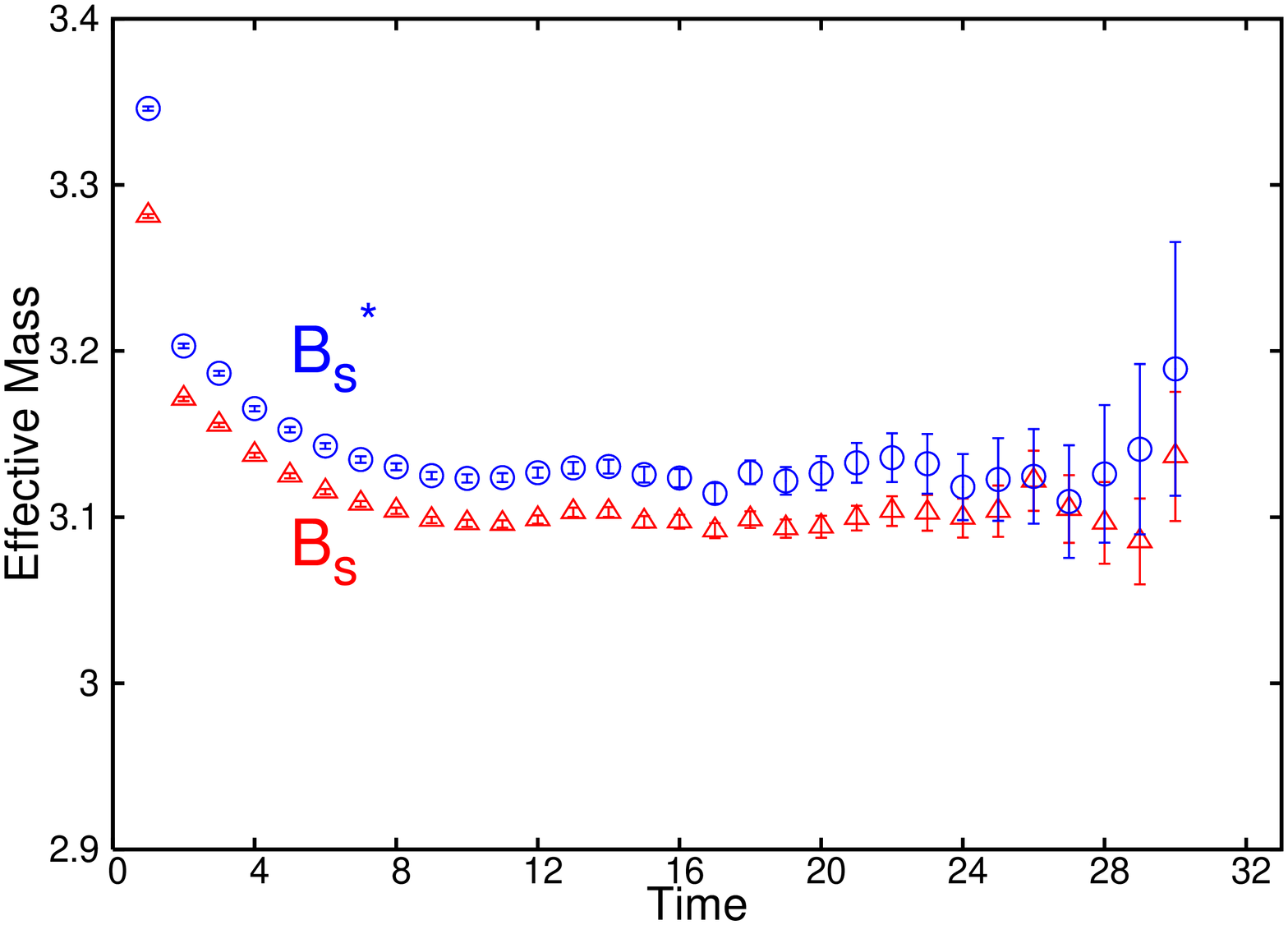,width=0.75\linewidth,angle=270}  
  \end{minipage}
  \caption{effective masses for the mesons on $m_{sea}=0.005$ ensembles: (left)$\eta_b$ and $\Upsilon$;
  (right)$B_s$ and $B_s^*$.}
  \label{fig:eff-mass-hUBsBss}
  \hfill
  \vspace*{\afterFigure}
\end{figure}

\begin{figure}[ht]
  \hfill
  \begin{minipage}{0.45\textwidth}
    \hspace{-1.0cm}
    \epsfig{file=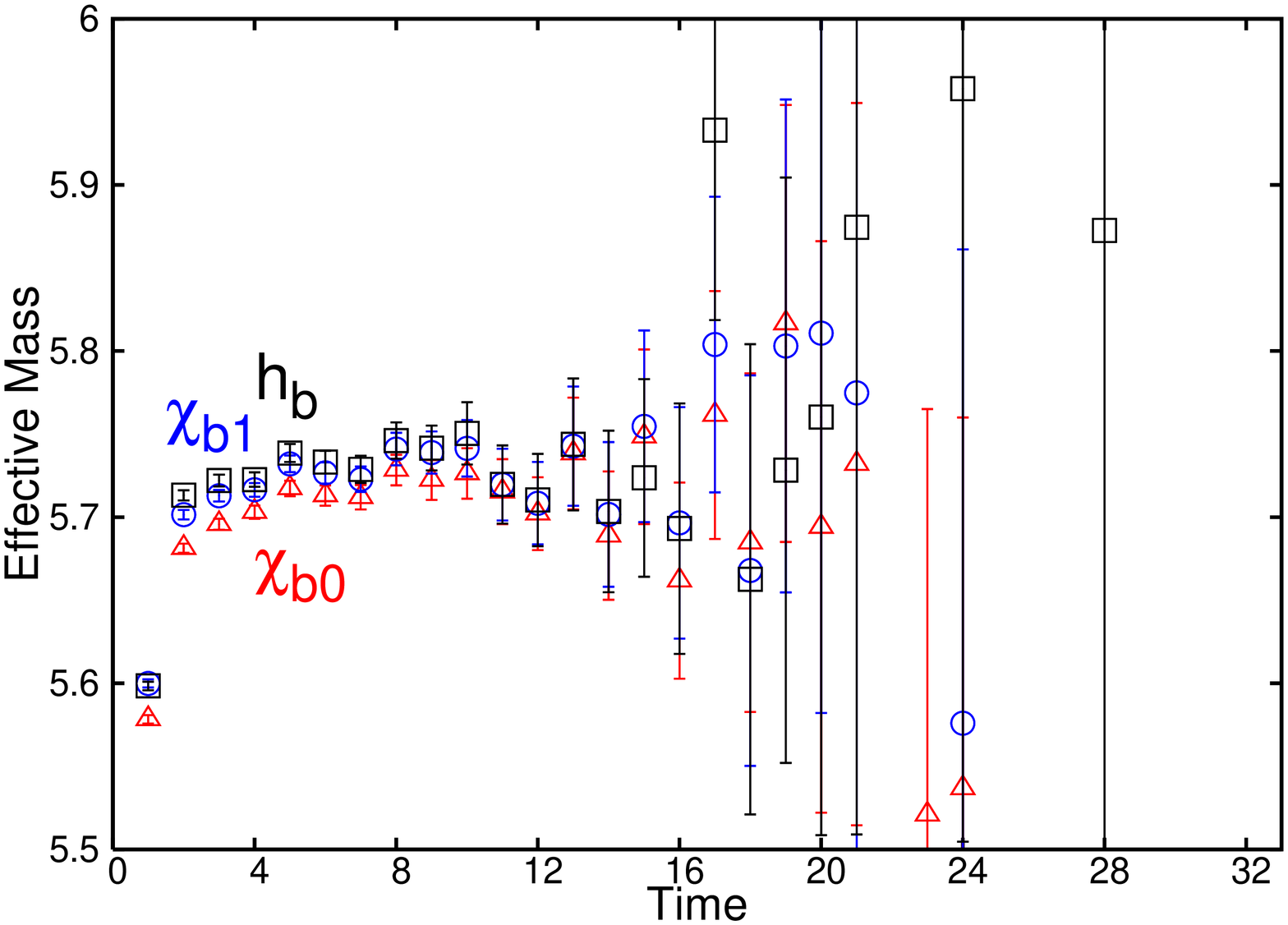,width=0.75\linewidth,angle=270}    
  \end{minipage}
  \hfill
  \begin{minipage}{0.45\textwidth}
    \hspace{-0.8cm}
    \epsfig{file=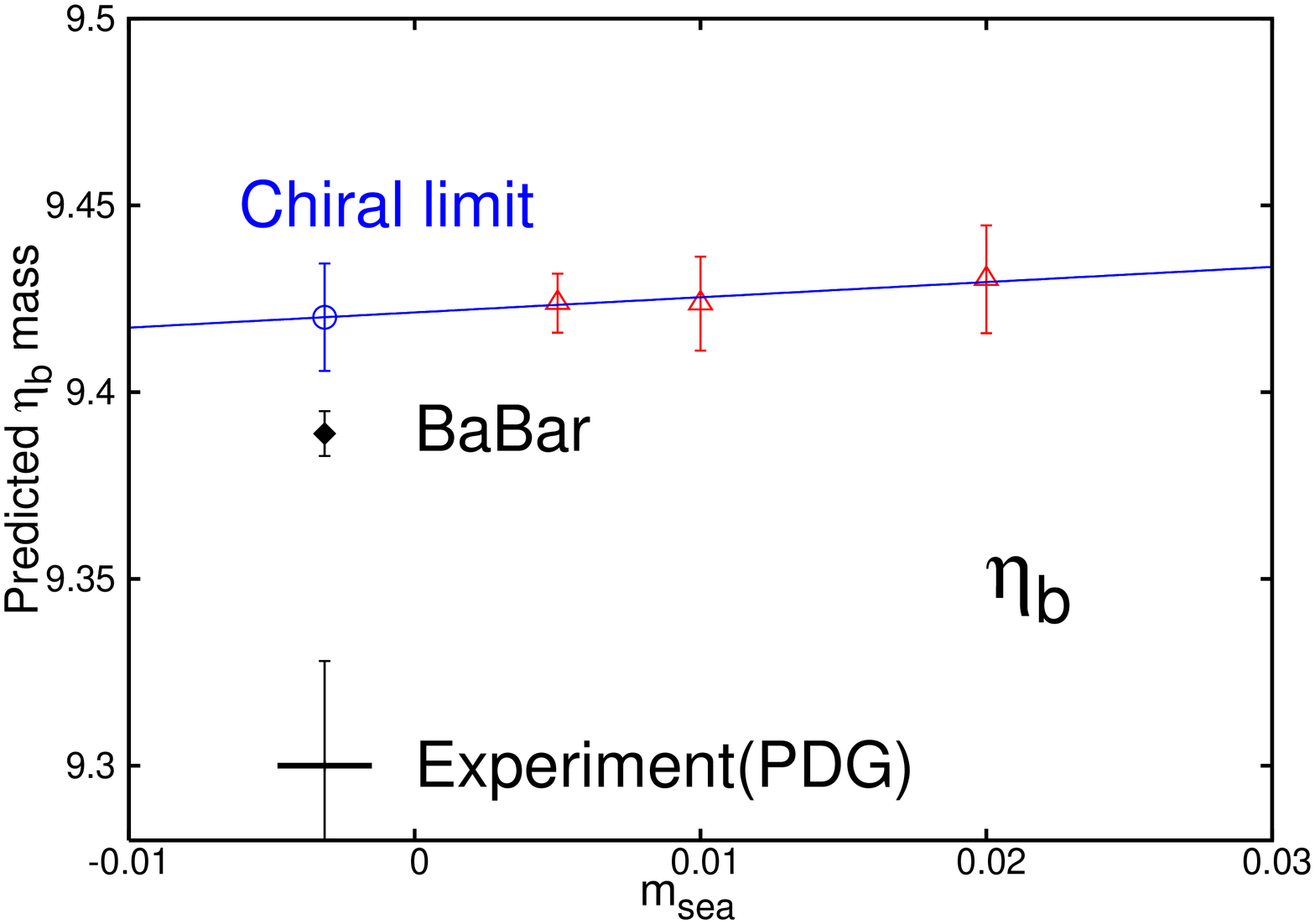,width=0.75\linewidth,angle=270}  
  \end{minipage}
  \caption{(left) Effective masses for the mesons: $\chi_{b0}$,$\chi_{b1}$ and $h_b$.
  (right) the prediction of $\eta_b$ mass in the chiral limit compared with the experiment.}
  \label{fig:eff-mass-chi-and-pred-etab}
  \hfill
  \vspace*{\afterFigure}
\end{figure}

For all the calculations we use a box source with size 4, which is not optimized to give the best plateau. The physical strange quark 
mass used in the bottom-strange spectrum is 0.034 in lattice units~\cite{Allton:2008pn}. Some sample effective mass plots for different states
are shown for the $m_{sea}=0.005$ ensemble, Fig~\ref{fig:eff-mass-hUBsBss} and ~\ref{fig:eff-mass-chi-and-pred-etab}(left). We can see from the 
plots that the plateau for $\chi_{b0}$,$\chi_{b1}$ and $h_b$ states are not quite clear so their masses are subject to more systematic errors. All meson correlators 
are fit to a single state with exponential decay. The fitting ranges are chosen from an examination of the plateau in the effective mass plots, 
and they are summarized below in Tab.~\ref{tab:fit-range}.
\begin{table}[ht]
  \centering
  \begin{tabular}{|c|c|c|c|}
    \hline
    fitting range & $\eta_b$/$\Upsilon$ & $\chi_{b0}$/$\chi_{b1}$/$h_b$ & $B_s$/$B_s^*$\\
    \hline
    $m_{sea}=0.005$ & [14,30]  & [5,12] & [10,25]   \\
    $m_{sea}=0.01$  & [11,30]  & [5,9]  & [10,25]   \\
    $m_{sea}=0.02$  & [11,30]  & [6,15] & [13,18]   \\
    \hline
  \end{tabular}
  \caption{The time fitting ranges for different mesons at different sea quark masses.}
  \label{tab:fit-range}
\end{table}

The momentum dependence is studied for both the $\eta_b$ and $\Upsilon$ mesons and the mass ratio $m_1/m_2$ extracted from each
are quite consistent. We use results from the $\Upsilon$ momentum dependence with the three lowest momenta. Please note
that although we are using bottom-strange spectrum to fix the RHQ parameters, the mass ratio $m_1/m_2$ is determined
from bottomonium because it is much better determined than it is from bottom-strange. To ensure the precision of the heavy 
quark propagator, we use an extremely 
tight CG stopping condition of $10^{-60}$. The CG stopping conditions are set to $10^{-8}$ for light quark propagators 
as before. 


\section{Analysis and Results}
\vspace*{\closersection}

Again we list all the quantities calculated in this work: (1)$\frac{m_1}{m_2}\;$(2)$m_{B_s^*}-m_{B_s}\;$(3)$\frac{1}{4}(m_{B_s}+3m_{B_s^*})\;$ (4)$m_{\Upsilon}-m_{\eta_b}\;$ (5)$\frac{1}{4}(m_{\eta_b}+3m_{\Upsilon})\;$ (6)$m_{\chi_{b1}}-m_{\chi_{b0}}\;$(7)$\frac{1}{4}(m_{\chi_{b0}}+3m_{\chi_{b1}})\;$(8)$m_{h_b}$

\subsection{Heavy-strange sector: the RHQ action parameters}
As we mentioned before, except for the mass ratio $m_1/m_2$ which is determined from the $\Upsilon$ momentum dependence, 
only the heavy-light, $B_s$ and $B_s^*$ masses are used to fix the RHQ action 
parameters for the bottom system. Throughout the analysis, the lattice scale is taken to be $a^{-1}=1.73$ GeV~\cite{Allton:2007hx}. 
From matching to quantities (1)-(3), the RHQ parameters and the corresponding chiral limit extrapolations are shown in Tab.~\ref{tab:rhq-para}.

\begin{table}[ht]
  \vspace{-0.3cm}
  \centering
  \begin{tabular}{|cccc|}
    \hline
    \hline
    $m_{sea}$ & $m_0a$ & $c_P$ & $\zeta$ \\ 
    \hline
    0.005 & 7.37(7)  & 3.84(40)  &  4.21(3)  \\
    \hline
    0.01  & 7.28(9)   & 3.28(40)  &  4.21(3)  \\
    \hline
    0.02  & 7.30(11)  & 3.52(53)  &  4.24(4)  \\
    \hline
    \hline
    -$m_{res}$& 7.38(12)  & 3.93(54)  & 4.19(4)\\
    \hline
    \hline 
  \end{tabular}
  \caption{The RHQ action parameters determined from matching the quantities (1)-(3), and extrapolated to the chiral limit.}
  \label{tab:rhq-para}
\end{table}

\subsection{Heavy-heavy sector: the predictions on bottomonium states}
What would we expect for the size of the errors on the bottomonium masses computed in this way? A naive estimate would be the size
of a typical order $a^2$ operators, for example, $\hat{O}=\overline{\Psi}\vec{\gamma}\cdot\vec{D}\;\vec{D}^2\Psi$.
The typical velocity in bottomonium system $v\sim 0.1$, which is determined by the splitting $\Upsilon(2S)-\Upsilon(1S) \sim mv^2 \sim 500$ MeV. This indicates $\langle\hat{O}\rangle\sim\frac{p^4a^2}{m_b}\sim 300$MeV.

Let's turn to the numerical results given by this work. All calculations are made from a complete calculation 
of each jackknife block from which the average and errors are determined. The results for various quantities are 
summarized in Tab.~\ref{tab:predictions} and compared to experimental values and other lattice calculations. Our 
results include only statistical errors. Experimental results are from PDG unless otherwise specified. We can see 
that most of our calculated bottomonium masses are within 30 MeV(<1\%) and one standard deviation of the experimental value. 
Of course, the results for the mass splittings give a more precise test: the spin-orbit splitting from our calculation is
about 2.5 standard deviations from the experiment and the hyperfine splitting is further away from the
value newly determined by BaBar collaboration~\cite{:2008vj} or lattice NRQCD calculation~\cite{Gray:2005ur}.
But the accuracy of our results indeed comes as a surprise! We were expecting errors of hundreds MeV but only see errors several
times smaller.

\begin{table}[ht]
  \vspace{-0.5cm}
  \centering
  \begin{tabular}{|c|c|c|c|}
    \hline
    \hline
    quantities & RHQ(MeV) & Exp.(MeV)  &  NRQCD(MeV) \\ 
    \hline
    $m_{\eta_b}$    & 9420(14)   &  9389(3)(3)~\cite{:2008vj} &  \\
    $m_{\Upsilon}$  & 9444(17)   &  9460     &  \\
    $m_{\chi_{b0}}$ & 9873(15)   &  9859     &  \\
    $m_{\chi_{b1}}$ & 9897(16)   &  9893     &  \\
    $m_{h_b}$       & 9908(17)   &   -       &  9900(6)~\cite{Gray:2005ur} \\
    $m_{\Upsilon}-m_{\eta_b}$ &  23.7(3.7)   &  71(3)(3)~\cite{:2008vj}  & 61(14)~\cite{Gray:2005ur} \\
    $m_{\chi_{b1}}-m_{\chi_{b0}}$ & 24.0(3.5)     & 33.34  &  \\
    \hline
    \hline
  \end{tabular}
  \label{tab:predictions}
  \caption{The predictions on individual masses and splittings, compared to experiment and lattice NRQCD
    calculations. Results from our calculation include statistical errors only.}
\end{table}
 
A possible better theoretical estimate would use a simple hydrogen-like Coulomb model to describe
bottomonium. In that case, we could define a $\Upsilon$ state as:
\begin{eqnarray}
  |\Upsilon, m_j\rangle &=& \int\frac{d^3\vec{p}_1d^3\vec{p}_2}{(2\pi)^{9/2}}\sum_{s_1s_2}\delta^{(3)}(\vec{P}-\vec{p}_1-\vec{p}_2)
  \phi(\frac{\vec{p}_1-\vec{p}_2}{2})_{1S} \langle1m_j|s_1s_2\rangle a^\dagger(\vec{p}_1,s_1)b^\dagger(\vec{p}_2,s_2)|0\rangle \nonumber
\end{eqnarray}
where $a$ and $b$ are free field quark and anti-quark annihilation operators as in
\begin{eqnarray}
  \Psi(\vec{x})=\int \frac{d^3\vec{p}}{\sqrt{2E_p}}\frac{1}{(2\pi)^3}\sum_s\{u^s(p)e^{i\vec{p}\cdot\vec{x}}a(\vec{p},s)+v^s(p)e^{-i\vec{p}\cdot\vec{x}} b^\dagger(\vec{p},s)\}
\end{eqnarray}
and $\phi(\vec{p})$ is the Fourier transform of the hydrogen atom wave function with mass $m/2$, the reduced mass of the 
bottomonium system.

The size of the matrix element of the operator $\hat{O}$ can then be estimated and yields the result:
\begin{eqnarray}    
  \langle\hat{O}\rangle \sim \frac{5}{8}m_b^3\alpha_s^4a^2  = 
  \begin{array}{l}
    \sim 40 MeV \;\;\quad m_b =4.0GeV, \alpha_s=0.25 \\
    \sim 146 MeV  \quad m_b=4.0GeV, \alpha_s=0.35
  \end{array} \nonumber 
\end{eqnarray}
Thus the size is very sensitive to the strong coupling constant $\alpha_s$. As an accurate $\alpha_s$ is not 
available we can not make any strong statement from this approach. 
\begin{figure}[ht]
  \hfill
  \begin{minipage}{0.45\textwidth}
    \hspace{-1.0cm}
    \epsfig{file=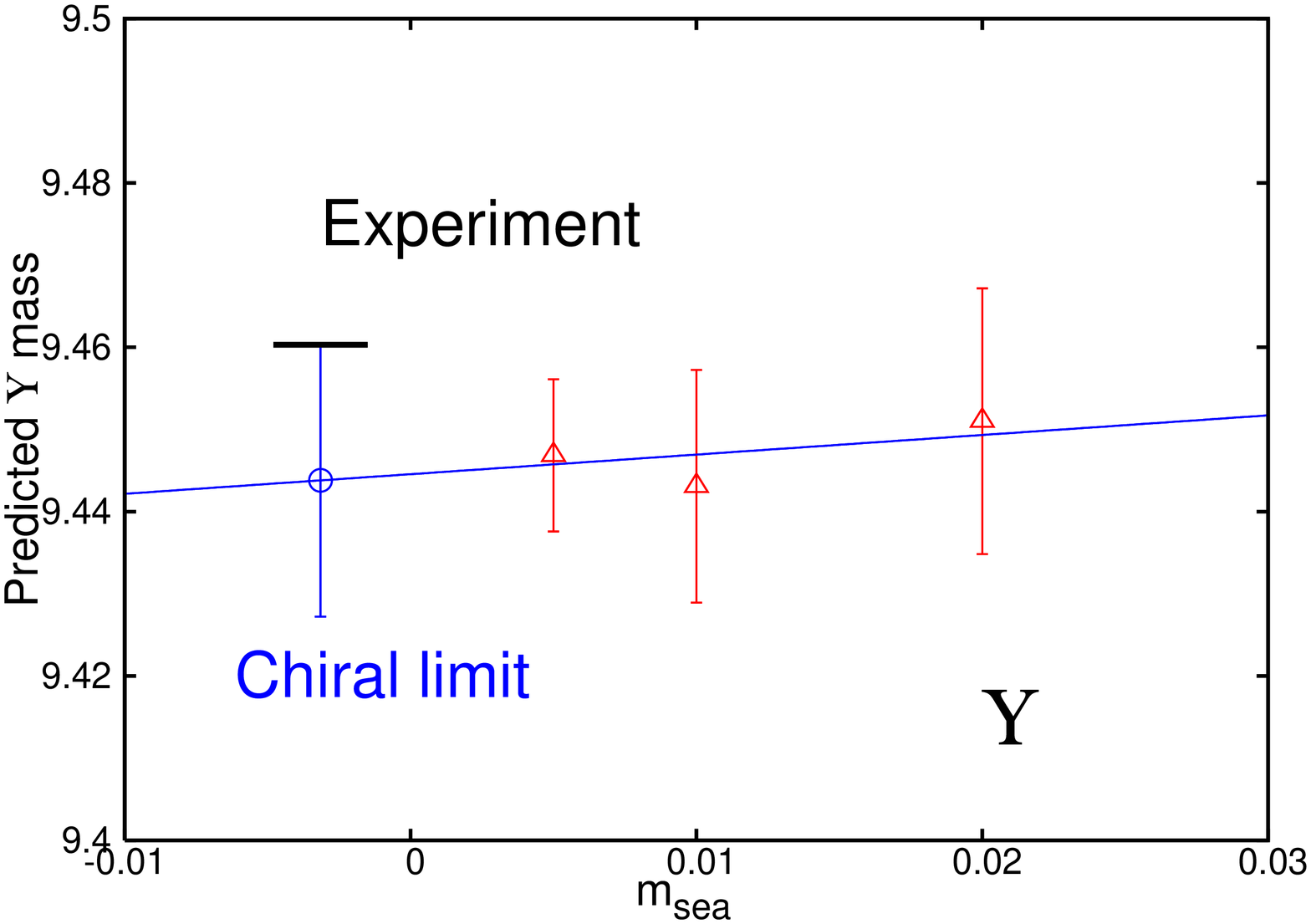,width=0.75\linewidth,angle=270}    
  \end{minipage}
  \hfill
  \begin{minipage}{0.45\textwidth}
    \hspace{-0.8cm}
    \epsfig{file=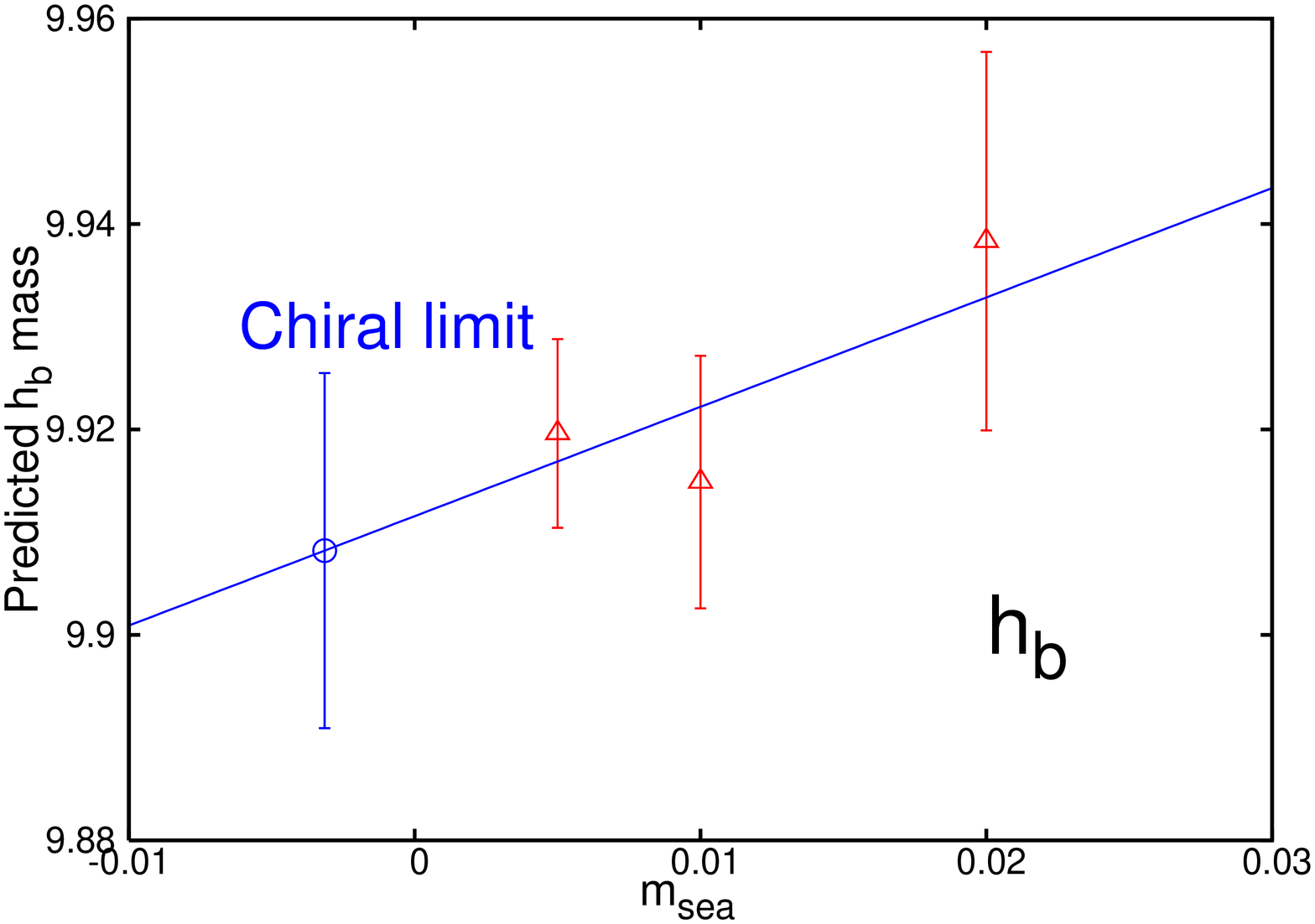,width=0.75\linewidth,angle=270}  
  \end{minipage}
  \caption{The mass predictions in the chiral limit compared with the experiment.
    (left) $\Upsilon$; (right) $h_b$.}
  \label{fig:pred-U-hb}
  \hfill
  \vspace*{\afterFigure}
\end{figure}
\begin{figure}[ht]
  \hfill
  \begin{minipage}{0.45\textwidth}
    \hspace{-1.0cm}
    \epsfig{file=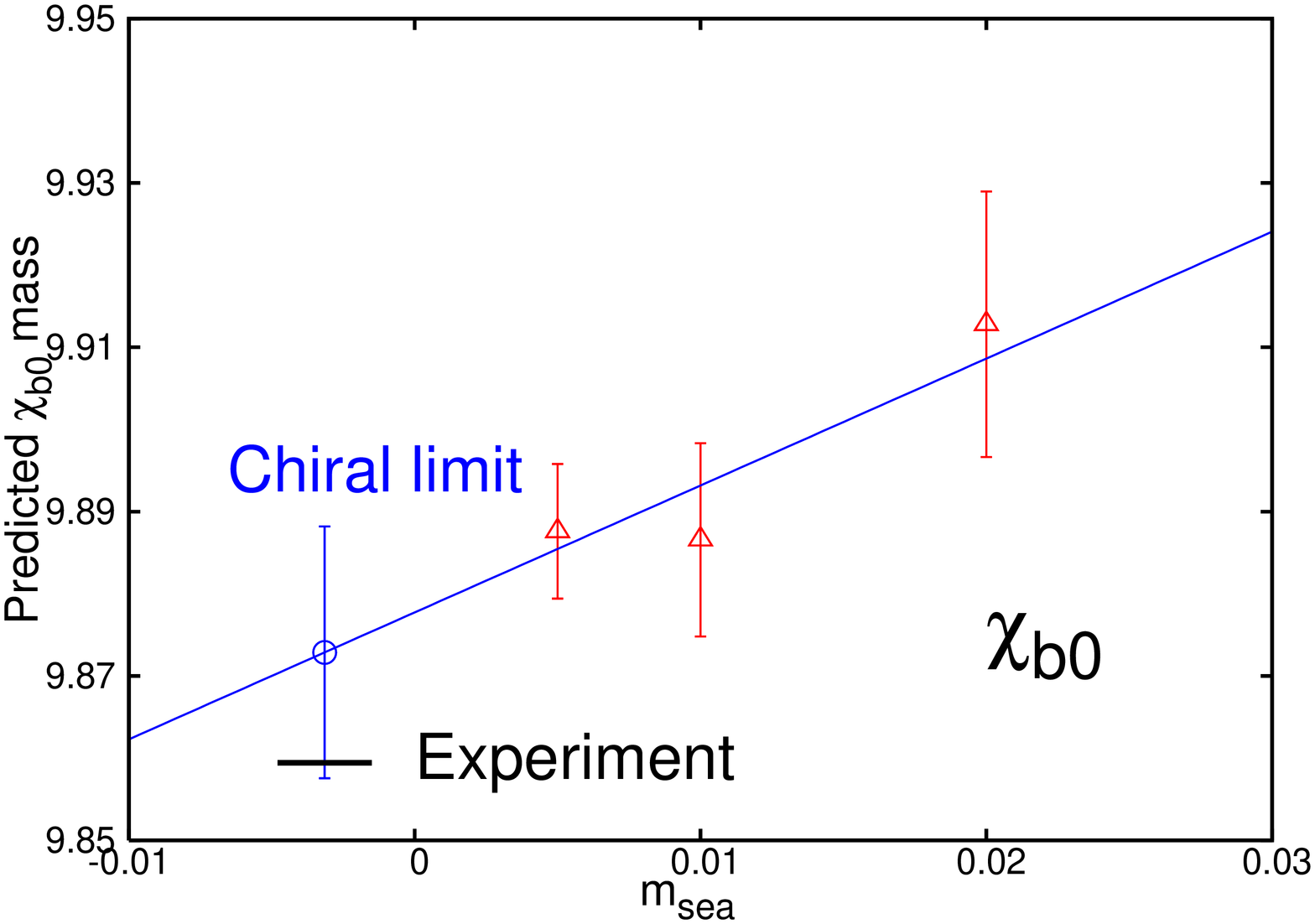,width=0.75\linewidth,angle=270}    
  \end{minipage}
  \hfill
  \begin{minipage}{0.45\textwidth}
    \hspace{-0.8cm}
    \epsfig{file=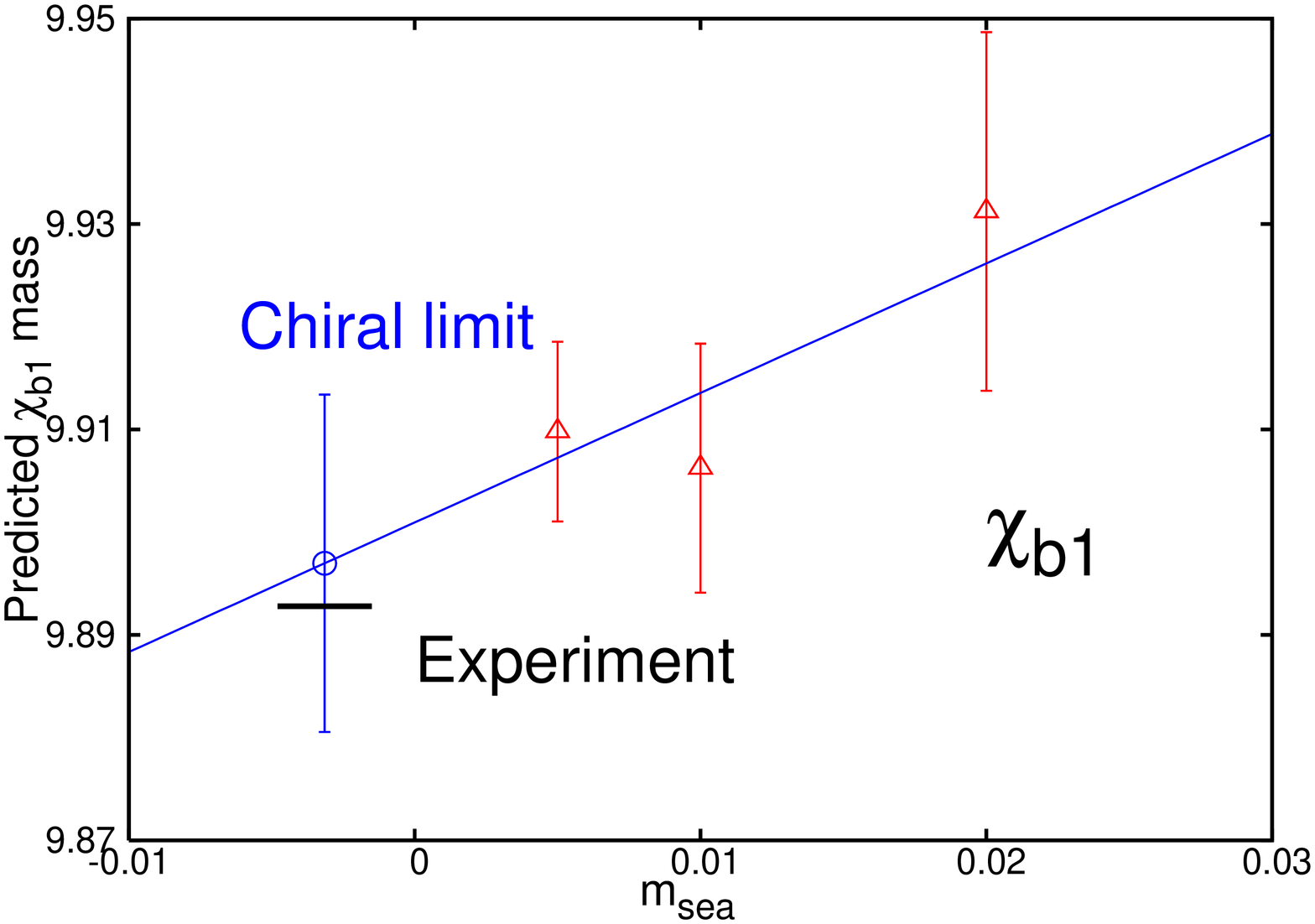,width=0.75\linewidth,angle=270}  
  \end{minipage}
  \caption{The mass predictions in the chiral limit compared with the experiment.
    (left) $\chi_{b0}$; (right) $\chi_{b1}$.}
  \label{fig:pred-chib01}
  \hfill
  \vspace*{\afterFigure}
\end{figure}

\section{Conclusion}
\vspace*{-0.15cm}

We have studied the bottomonium and bottom-strange system. We calculate the RHQ parameters by matching the bottom-strange system 
to the experiment at each sea quark mass and extrapolate to the chiral limit. The bottomonium spectrum is then determined in
the chiral limit within 30 MeV of the experimental values. A naive
estimate of the size of a $a^2\vec{p}^2$ operator suggests the error expected in bottomonium spectrum should be a few hundred MeV, 
while a more careful hydrogen-like Coulomb model suggests the error size is very sensitive to the strong coupling constant $\alpha_s$. 
Other phenomenological models might be useful to estimate the error size more accurately. In summary, our application
of the RHQ action on the bottom system yields surprisingly accurate results. Our possible next steps include study of heavy-light
system which could double-check the validity of the RHQ method in this regime, more calculations on other states like $\bar{b}c$ system
and nucleons with one or more heavy quarks, and matrix elements. These calculations could also be repeated using the new $32^3\times 64$ 
lattices to better determine the continuum limit.

\section*{Acknowledgment}
\vspace*{\closersection}
We acknowledge helpful discussions with Norman Christ, Robert Mawhinney and Huey-Wen Lin. This work was performed on the QCDOC computers
at BNL, Columbia, Edinburgh and the RBRC at BNL, and was supported by U.S. DOE grant DE-FG02-92ER40699.

\bibliographystyle{apsrev}
\bibliography{Proceeding}

\end{document}